\documentclass[onecolumn,showpacs,preprintnumbers]{revtex4}
\usepackage{pifont}
\usepackage{subfigure}
\usepackage{epsfig}
\usepackage{CJK}
\usepackage{amsmath}
\usepackage{amsfonts}
\usepackage{amssymb}
\usepackage{color}
\usepackage{graphicx}
\usepackage{mathrsfs}
\usepackage[normalem]{ulem}
\usepackage{overpic}
\usepackage{xcolor}

\newcommand{\bea}{\begin{eqnarray}}
\newcommand{\eea}{\end{eqnarray}}
\newcommand\btd{\raise 2pt
\hbox{$\hat\bigtriangledown$}\hskip 1.5pt}
\newcommand\bt{\raise 2pt
\hbox{$\bigtriangledown$}\hskip 1.5pt}

\newcommand{\omits}[1]{}

\UseRawInputEncoding
\begin{document}
\title{Photon Spheres and Shadows of Black Holes with Modified Entropy-Inspired Metrics}
\author{Fang Liu$^{a,b}$, Yu-Bo Ma$^{a,b}$, Yun-Zhi Du$^{a,b}$, Huai-Fan Li$^{b,c}$}
\thanks{Corresponding author.\\ E-mail address: fangliuphys@sxdtdx.edu.cn (Fang Liu), yuboma.phy@gmail.com(Yu-Bo Ma), duyzh13@lzu.edu.cn (Yun-Zhi Du), huaifan.li@stu.xjtu.edu.cn (Huai-Fan Li).\\
Supported by the National Natural Science Foundation of China (Grant No. 12375050), the Natural Science Foundation of Shanxi Province (Grant No. 202303021211180, 202203021221209, 202203021221211, 202503021211241), the Program for the Innovative Talents of Higher Education Institutions of Shanxi (Grant No. 2024Q030), and the Doctoral Sustentation Fund of Shanxi Datong University.}

\affiliation{{\footnotesize $^a$ Department of Physics, Shanxi Datong University, Datong 037009, China}\\
{\footnotesize $^b$ Insitute of Theoretical Physics, Shanxi Datong University, Datong 037009,  China}\\
{\footnotesize $^c$ College of General Education, Shanxi College of Technology, Shuozhou 036000, China}}

\begin{abstract}
Starting from the first law of black hole thermodynamics, we establish an explicit correspondence between the corrected entropy and the metric function under the condition of fixed black hole energy and horizon position. Using the corrected metric, we further compute the photon sphere radius and shadow size, demonstrating that different entropy corrections lead to characteristic optical shifts. By comparing with the Event Horizon Telescope observations of Sgr A*, we constrain the parameter range introduced in the corrected entropy. This provides a feasible approach for testing generalized entropy frameworks and probing deviations from the Bekenstein-Hawking area law.
\end{abstract}

\maketitle





\section{Introduction}
The connections among gravity, thermodynamics, and quantum theory have proven highly useful in studying the structure of spacetime. Since the pioneering work of Bekenstein and Hawking revealed that black holes behave as thermodynamic objects with well-defined temperature and entropy, it has become evident that the laws of gravity are intimately linked to thermodynamic structure \cite{SWH-1975,JDB-1973,RGC-2016,DK-2012,RGC-2013,JLZ-2015}.

One of the most profound predictions of general relativity is the deflection of light in curved spacetimes, known as the gravitational lensing effect. Gravitational lensing has emerged as a crucial tool for addressing fundamental problems in astrophysics and cosmology. Research on gravitational lensing in the strong$-$field regime, particularly near compact objects such as black holes and neutron stars, began in the 1970s \cite{JP-1979}. Since then, extensive studies have explored gravitational lensing caused by a wide range of structures, including dark matter, dark energy, quasars, gravitational waves, and other compact objects \cite{FA-2021,MB-2006,SC-2015,KL-2016,NT-2022}.

Near black holes, (the most mysterious objects in the universe), light is not merely swallowed, it may also be temporarily trapped. Theoretically, a special spherical surface known as the photon sphere exists just outside the black hole event horizon, where gravity is strong enough to force photons to move along closed orbits, forming a short-lived and highly unstable ``photon cage". Any tiny perturbation will cause photons to either fall into the event horizon or escape to infinity. On April 10, 2019, astronomers around the world jointly released the first image of a black hole \cite{KA-2019,KA-2024a,KA-2024b}, and the photon sphere serves as the key structure underlying these images. The optical detection of the supermassive black hole at the galactic center Sgr A* (represented by the first black hole shadow image released by the Event Horizon Telescope (EHT)) stands as a landmark triumph for general relativity. Its far-reaching theoretical significance, however, lies in the following: for the first time, humanity's exploration of gravity has transcended the weak-field approximation and entered an entirely new regime encompassing strong gravitational fields, dynamical spacetime, and quantum effects, while also furnishing a practical instrument capable of stringently constraining a broad spectrum of new physics theories.
In essence, the optical imaging of Sgr A* marks a revolutionary leap in gravitational testing methodologies: from ``hearing" gravity via gravitational waves to ``seeing" it through electromagnetic radiation. This breakthrough ushers strong-gravity astronomy¡¯s observational and theoretical investigations into an unprecedented era of high-precision research. In particular, the black hole shadow observed in these images is closely related to the properties of the photon sphere and the strong gravitational lensing effect near the event horizon. In recent years, the study of photon spheres and black hole shadows has attracted considerable interest, with a large number of works investigating the photon spheres and shadow images of various black hole models. On the other hand, photon spheres can also provide valuable insights into the physical properties and characteristics of black holes. They can reveal nontrivial features related to spacetime singularities, event horizons, causal structures, and asymptotic behaviors associated with black holes and ultracompact objects. Significant differences exist in the size and morphology of shadows cast by different types of black holes, offering important clues for probing black hole properties and serving as a powerful tool for estimating black hole parameters and testing general relativity as well as alternative modified theories of gravity \cite{JB-2026,SC-2024,SDB-2025,SC-2025,AA-2026a,JML-2026,CYY-2026,EB-2026,CKQ-2024,XXZ-2026a,XXZ-2026b,XXZ-2025a,XXZ-2025b,XXZ-2025c,BFW-2025,XXZ-2020,XXZ-2025d,
JW-2026,TYZ-2026,RB-2026,SH-2026,SWW-2024,SBC-2023,YYW-2026,CB-2019,MK-2024,PKY-2025,JLC-2025,WTL-2025,AE-2025a,KJH-2025a,SR-2024,MIA-2024,HL-2024,AS-2026,ZYL-2026,YZD-2023,YZD-2025,JS-2026,FA-2026a,CP-2020,SNG-2025,UZ-2026}.

Significant progress has also been made in understanding photon spheres and black hole shadows from a geometric perspective. For static, spherically symmetric spacetimes, the photon sphere can be characterized by the vanishing of geodesic curvature and Gaussian optical curvature, providing a coordinate-invariant framework for investigating circular null geodesics.
Recently, the authors of \cite{KJH-2026} systematically studied the gravitational lensing effect and dynamical evolution of the shadow of the Vaidya black hole using backward ray-tracing techniques, paving the way for exploring the optical properties of nonstationary black holes. The authors of \cite{CKQ-2024} investigated the relation between the number of stable and unstable photon spheres outside the black hole horizon. Motivated by these developments, recent studies on the optical properties of various deformed black holes have attracted our attention \cite{HEG-2025,MFF-2026,RCDP-2025,FA-2026b,CFSP-2026,MAR-2026,HLZ-2026}. The relationships among the location of the black hole photon sphere, critical impact parameters, and the effective potential are among the key issues of widespread concern. Compared with Schwarzschild black holes, black holes with quantum corrections generally possess larger radii for the event horizon, photon sphere and photon ring, alongside a lower effective potential for photons. This conjecture has been widely accepted in the community \cite{JP-2021,HL-2020}. In contrast, typical black holes within nonlinear electrodynamics exhibit opposite behaviors relative to Schwarzschild black holes: their event horizon, photon sphere and photon ring radii are smaller, while the photon effective potential becomes higher \cite{SG-2023,XXZ-2022}. Our analysis further reveals that distinct conclusions can be drawn for different entropy corrections.

A central goal of modern theoretical physics is to understand the microscopic origin of black hole entropy. Since Bekenstein and Hawking proposed that black hole entropy is proportional to the area of its event horizon, the concept of corrected Bekenstein$-$Hawking entropy attracted extensive research attention \cite{SN-2021,SN-2022,EE-2025,HXL-2025}. The corrected Bekenstein$-$Hawking entropy serves as an important bridge connecting general relativity with quantum gravity theories. From loop quantum gravity to string theory, diverse theoretical frameworks predict correction terms to the entropy, including logarithmic corrections, nonextensive entropies (such as R$\acute{e}$nyi entropy and Sharma$-$Mittal entropy), and Barrow entropy that accounts for quantum gravity fluctuations. Any modification to the Bekenstein$-$Hawking entropy$-$area relation induces a corresponding deviation in the spacetime metric, which in turn directly alters the position, stability, and even topological properties of the photon sphere. Recently, the authors of \cite{AA-2026b,AA-2026c} demonstrated that, starting from a chosen corrected entropy functional, one can systematically derive the associated modified metric and its effective matter sector, thereby establishing a direct entropy$-$geometry correspondence. Furthermore, by comparing with Event Horizon Telescope observations of Sgr A* at the Galactic center, quantitative constraints were imposed on the parameters introduced in the corrected entropy.
Very recently, the authors of \cite{AA-2025} derived the van der Waals black hole by imposing the condition that the temperature of the van der Waals system equals the Hawking radiation temperature of the black hole. Inspired by this idea, and based on the first law of thermodynamics satisfied by any thermodynamic system, we investigate the effects of different generalized entropies on the spacetime structure of black holes. First, we apply the first law of thermodynamics to obtain the black hole radiation temperature corresponding to various entropy corrections. It should be pointed out that under the same spacetime background, the event horizon of the black hole ought to remain unchanged for different entropy corrections. For the Schwarzschild black hole $r_+=2M$ is held fixed. Under these two constraints, we construct the relation between the corrected entropy and the spacetime metric. We obtain the metric corresponding to the corrected entropy that preserves the fixed horizon position $r_+$ and $r_+=2M$, as well as the black hole radiation temperature $T={{\left. \frac{f'(r,\alpha ,\beta \cdots )}{4\pi } \right|}_{r={{r}_{+}}}}$, where $T$ denotes the temperature at which spacetime thermodynamic quantities satisfy the first law of thermodynamics after entropy correction.

The structure of this paper is organized as follows: In Section \ref{nonextensiveentroy}, we briefly review various entropy corrections. In Section \ref{processes}, using the metric consistent with the corrected entropy, we investigate the photon sphere location and shadow radius of the modified black hole, analyze the influence of each parameter on the photon sphere and black hole shadow, and compare the theoretical results with the observational data of Sgr A* at the Galactic center to obtain constraints on the corresponding parameters. Concluding remarks and a detailed discussion of the findings are presented in Section \ref{conslusion}. we use the units $G_d= \hbar=k_B=c=1$ throughout this paper.

\section{Non-extensive Entropy}\label{nonextensiveentroy}

Introduced by Constantino Tsallis, non-extensive entropy serves as an extension of the conventional Boltzmann-Gibbs entropy. This notion is particularly valuable for systems characterized by non-linearity and a strong sensitivity to initial conditions. In contrast to the Boltzmann-Gibbs entropy, which presupposes a linear scaling of entropy with the system size, non-extensive entropy is capable of accommodating systems where such linearity breaks down. As a result, it finds applications across a broad spectrum of fields, including theoretical physics, cosmology, and statistical mechanics. It is of particular relevance to systems featuring long-range interactions, fractal structures, or memory effects \cite{AA-2024,SNG-2025}.
\subsection{Barrow Entropy}
Barrow proposed that quantum gravitational corrections could alter the classical smoothness of the event horizon, leading to a horizon geometry with fractal features. This modification suggests that the standard area law governing black hole entropy does not hold precisely. To quantify the level of this geometric irregularity, a parameter $\Delta$ is introduced, which characterizes the extent to which the horizon deviates from a smooth two-dimensional surface. Incorporating this modification, the entropy associated with a black hole can be expressed as \cite{SNG-2025b,JDB-2020}:
\begin{equation}
{{S}_{B}}={{({{S}_{BH}})}^{1+\tfrac{\Delta }{2}}},~~~~~~{{S}_{BH}}=S_{B}^{\tfrac{2}{2+\Delta }},\label{Sbb}
\end{equation}
where the parameter satisfies $0 \leq \Delta \leq 1$. When $\Delta = 0$, the horizon remains undeformed, and the expression reduces to the standard Bekenstein-Hawking entropy; in contrast, non-zero values of $\Delta$ encapsulate the effects of the fractal structure induced by quantum gravity. Conceptually, a non-zero $\Delta$ signifies that the microstructure of spacetime at the event horizon departs from classical smoothness, which may reflect the underlying quantum gravitational degrees of freedom. As such, Barrow entropy offers an effective approach to modeling these quantum gravitational corrections without requiring a detailed specification of the microscopic theory.
\subsection{R$\acute{e}$nyi Entropy}
R$\acute{e}$nyi Entropy is also one form of non-extensive entropy that has been employed in the study of black hole thermodynamics \cite{AR-1961,SNG-2025,CP-2020}. It provides a generalized entropy measure that goes beyond the additive framework of the Bekenstein-Hawking entropy. The key feature of this framework is the parameter $\lambda$, which  quantifies the deviation of the entropy from extensivity. This modified entropy construction proves valuable for black hole thermodynamics, as the microscopic horizon degrees of freedom can exhibit interactions unaccounted for within conventional Boltzmann-Gibbs statistics. The R$\acute{e}$nyi entropy associated with a black hole can be expressed as

\begin{equation}
{{S}_{R}}=\frac{1}{\lambda }\ln (1+\lambda {{S}_{BH}}),~~~~~~{{S}_{BH}}=\frac{{{e}^{\lambda {{S}_{R}}}}-1}{\lambda }.\label{Srb}
\end{equation}

The parameter $\lambda$ in non-extensive entropy plays a pivotal role in defining the entropy function. In the expression, the factor $(1+\lambda {{S}_{BH}})$ must be a dimensionless quantity. The the physical dimension of ${S}_{BH}$ is $[L^{2}]$, it follows that the physical dimension of $\lambda$ is $[L^{-2}]$.  From the perspective of thermodynamics and statistical mechanics, the parameter $\lambda$ in this entropy should be $0 < \lambda < 1$ \cite{CP-2020}.
\subsection{Sharma$-$Mittal Entropy}
Sharma-Mittal entropy constitutes another important class of nonextensive entropy, which unifies and generalizes both R$\acute{e}$nyi and Tsallis entropies. This entropy has found prominent applications in cosmology; for instance, it can account for the cosmic accelerated expansion via an effective description of vacuum energy. While various nonextensive entropies have been widely adopted for black hole research, Sharma-Mittal entropy remains underexplored within black hole thermodynamic contexts. This motivates us to investigate black hole thermodynamic characteristics from the perspective of Sharma-Mittal entropy, treating black holes as strongly coupled gravitational systems \cite{AM-2017,ASJ-2018,MASA-2025}.

It can be express as fowllow:
\begin{equation}
{{S}_{SM}}=\frac{1}{\alpha }\left( {{(1+\beta {{S}_{T}})}^{\tfrac{\alpha }{\beta }}}-1 \right),~~~~~~d{{S}_{SM}}={{(1+\beta {{S}_{BH}})}^{\tfrac{\alpha }{\beta }-1}}d{{S}_{BH}},~~~~~~\frac{\partial {{S}_{BH}}}{\partial {{S}_{SM}}}={{(1+\alpha {{S}_{SM}})}^{1-\tfrac{\beta }{\alpha }}},\label{Sbs}
\end{equation}
In this expression, ${S}_{T}=\frac{A}{4}$, where $A=4\pi r^2_+$ and $r_+$ is the horizon radius, represents the Tsallis entropy. The parameters $\alpha$ and $\beta$ are free and must be determined by fitting the results with observational data. We notice that factors $(1+\beta {{S}_{T}})$ and $(1+\alpha {{S}_{SM}})$ are dimensionless, accordingly, the dimensions of  $\alpha$ and $\beta $  are both $[L^{-2}]$. Notably, when $\alpha \to 0$ and $\alpha \to \beta $, Sharma-Mittal entropy reduces to R$\acute{e}$nyi and Tsallis entropies, respectively \cite{SNG-2024bbb}.
\section{Modified spacetime geometry from modified horizon entropy}\label{processes}
 In classical general relativity, the standard procedure involves specifying an energy-momentum tensor, solving for the metric components, and then computing the corresponding horizon entropy. Conversely, concepts such as emergent gravity and the holographic principle suggest that gravity may originate from entropic or informational degrees of freedom, highlighting the fundamental role of entropy. Guided by this perspective, we propose a straightforward approach wherein the spacetime geometry is derived directly from the horizon entropy.

 The general static spherically symmetric line element is given by
 \begin{equation}
 ds=-f(r)d{{t}^{2}}+{{f}^{-1}}(r)d{{r}^{2}}+{{r}^{2}}(d{{\theta }^{2}}+{{\sin }^{2}}\theta d{{\varphi }^{2}}),\label{ds}
\end{equation}
The event horizon $r_+$ satisfies $f(r_+)=0$,  and the Hawking temperature can be obtained from ${{T}_{BH}}=f'({{r}_{+}})/4\pi $. Meanwhile, the first law of black hole thermodynamics is written as:
\begin{equation}
dM={{T}_{BH}}d{{S}_{BH}}.
\end{equation}
For a static spherically symmetric spacetime, when the Bekenstein-Hawking entropy $S_{BH}$ of the black hole is corrected to $S$, the state parameters of the black hole must satisfy the first law of thermodynamics, namely
\begin{equation}
T=\frac{\kappa }{2\pi }=\frac{f'({{r}_{+}})}{4\pi }=\frac{\partial M}{\partial S}=\frac{\partial M}{\partial {{S}_{BH}}}\frac{\partial {{S}_{BH}}}{\partial S}={{T}_{BH}}\frac{\partial {{S}_{BH}}}{\partial S}.\label{T1}
\end{equation}
Note that the parameters introduced via entropy corrections are treated as constants.
How corrected entropy affects the spacetime geometry of black holes is a question of widespread concern. Recently, the authors of \cite{AA-2026b,AA-2026c} proposed the entropy-geometry correspondence scheme to address the modification of the spacetime metric induced by corrected entropy. A key aspect of this approach is replacing the coordinate function $r$ in the Schwarzschild metric with $S$, thereby obtaining the influence of entropy corrections on the metric and the resulting photon sphere. The authors of \cite{AA-2025} proposed applying the condition that the temperature of the van der Waals system is equal to the Hawking temperature of the black hole, yielding the spacetime metric corresponding to the van der Waals black hole. This idea has opened up a new pathway for our investigation of metric modifications induced by corrected entropy. The correction of black hole entropy originates from attempts to incorporate the effects of quantum gravity theories into classical black hole thermodynamics. Regardless of the method used to obtain the entropy correction terms, the core constraints are maintaining the black hole¡¯s parameter $M$ as a constant, keeping the position of the black hole¡¯s event horizon unchanged, and ensuring that the black hole¡¯s thermodynamic quantities satisfy the first law of thermodynamics. Now we discuss how to derive the spacetime metric corresponding to entropy corrections under these constraints.

We assume that the backreaction induced by any arbitrary entropy function can be reflected in the spacetime metric, that is, modifying the Schwarzschild metric Eq.(\ref{ds}) to
\begin{equation}
ds=-f(r,\alpha,\beta \cdots )d{{t}^{2}}+{{f}^{-1}}(r,\alpha,\beta \cdots )d{{r}^{2}}+{{r}^{2}}(d{{\theta }^{2}}+{{\sin }^{2}}\theta d{{\varphi }^{2}}),\label{ds1}
\end{equation}
where $\alpha,\beta,\cdots $ denotes the parameters introduced in different entropy corrections. These parameters are required to satisfy

\begin{equation}
{{\left.\frac{1}{4\pi}\frac{\partial f(r,\alpha,\beta,\cdots )}{\partial r}\right|}_{r={{r}_{+}}}}=T,\label{frat}
\end{equation}
where $T$ is given by Eq.(\ref{T1}). Secondly, the position of the black hole event horizon $r_+$  must satisfy $f({{r}_{+}},\alpha,\beta,\cdots )=0$  under the premise that $r_+=2M$ remains unchanged. Under these constraints, we take
\begin{equation}
f(r,\alpha \text{,}\beta \text{,}\cdots )=\frac{\partial {{S}_{BH}}}{\partial S}f(r),\label{fr}
\end{equation}
where $f(r)=1-\frac{2M}{r}$.  Since the photon sphere is located outside the horizon, typically around $r\sim 3M$ , in the Schwarzschild case, its position and the shadow radius are sensitive to the off-horizon extension of the metric. One can add corrections satisfying
\begin{equation}
\delta f({{r}_{+}})=0,~~~~~~~~\delta f'({{r}_{+}})=0. \label{fr+}
\end{equation}
Hence, Eq. (\ref{fr}) indicates the corrected black hole event horizon depends on $\frac{\partial {{S}_{BH}}}{\partial S}=0$ or $f(r)$. From Eqs. (\ref{Sbb}), (\ref{Srb}) and (\ref{Sbs}), we know that $\frac{\partial {{S}_{BH}}}{\partial S}\neq0$, which preserve the same horizon position but change the radius of the photon sphere. Furthermore, Eq.(\ref{fr}) must reduce to the Schwarzschild black hole case when all parameters $\alpha,\beta,\cdots $ are set to zero. To investigate the influence of entropy on the metric, we adopt the method proposed in Refs. \cite{AA-2026b,AA-2026c} for extending the black hole horizon relation, i.e., extending the Bekenstein-Hawking entropy ${{S}_{BH}}=\pi r_{+}^{2}$ to ${{S}_{BH}}\to \pi {{r}^{2}}$ and extending $r_+$ in various corrected entropies to $r$.

Thus the components of the Einstein tensor $G_{\nu }^{\mu }$ can be calculated by the metric (\ref{fr}). As we know from Ref. \cite{AE-2025b}
\begin{equation}
G_{t}^{t}=G_{r}^{r}=\frac{1}{r}\frac{df(r,\alpha ,\beta,\cdots )}{dr}+\frac{f(r,\alpha,\beta,\cdots )-1}{{{r}^{2}}},\label{gtt}
\end{equation}
\begin{equation}
G_{\theta }^{\theta }=\frac{{{d}^{2}}f(r,\alpha,\beta,\cdots )}{2d{{r}^{2}}}+\frac{1}{r}\frac{df(r,\alpha,\beta,\cdots )}{dr}.\label{gbb}
\end{equation}
From the field equations of general relativity
\begin{equation}
G_{\nu }^{\mu }=8\pi T_{\nu }^{\mu }, \label{gmm}
\end{equation}
we conclude that a non-zero effective stress-energy tensor originating from the Extended Generalized Uncertainty Principle (EGUP) can be obtained. In particular, we can define $T_{\nu }^{\mu }=(-\rho ,{{p}_{r}},{{p}_{t}},{{p}_{t}})$ , where
\begin{equation}
\rho =-G_{t}^{t},~~~\rho =-{{p}_{r}},~~~{{p}_{t}}=G_{\theta }^{\theta }. \label{pppt}
\end{equation}
It should be noted that the relation  reflects a radial equation of state similar to that of vacuum or dark energy \cite{MFF-2026,AA-2026b,AA-2026c}. However, due to ${{p}_{r}}\ne {{p}_{t}}$ , the effective matter sector originating from entropy is anisotropic. Consequently, the deviation of the entropy function from the standard area law acts as a geometric source, generating an anisotropic stress-energy tensor without the need to introduce any explicit matter fields. In this sense, the corrected entropy manifests as an effective, emergent gravitational matter content associated with the microscopic structure of the horizon. This conclusion is consistent with the discussions presented in Refs. \cite{AA-2026b,AA-2026c} as well as the results obtained when considering the generalized uncertainty principle \cite{HLZ-2026}.

The motion of massless photons along their geodesics is governed by the Lagrangian density \cite{CKQ-2022a,CKQ-2022b,JLC-2022,WTL-2025}
\begin{equation}
H=\frac{1}{2}{{g}^{\mu \nu }}{{p}_{\mu }}{{p}_{\nu }}=0,\label{hh}
\end{equation}
where ${{p}_{\mu }}=\frac{d{{x}_{\mu }}}{d\lambda }$ denotes the generalized momentum and $\lambda$ is the affine parameter. Owing to the spherical symmetry of this system, we focus on equatorial geodesics with $\theta =\pi /2$ , such that the Lagrangian simplifies to
\begin{equation}
-f(r,\alpha ,\beta \text{,}\cdots ){{\left( \frac{dt}{d\tau } \right)}^{2}}+{{f}^{-1}}f(r,\alpha ,\beta \text{,}\cdots ){{\left( \frac{dr}{d\tau } \right)}^{2}}+{{r}^{2}}{{\left( \frac{d\varphi }{d\tau } \right)}^{2}}=0,\label{ll}
\end{equation}
where $\tau$ is the proper time. Furthermore, the static and spherically symmetric spacetime possesses two conserved quantities, given by
\begin{equation}
E=-\frac{\partial H}{\partial \dot{t}}=f(r,\alpha ,\beta \text{,}\cdots )\dot{t},~~~~ \bar{L}=-\frac{\partial H}{\partial \dot{\varphi }}={{r}^{2}}\dot{\varphi }. \label{el}
\end{equation}
Using $d{{s}^{2}}=0$, the radial component of the geodesic equations can be rewritten as

\begin{equation}
 \frac{1}{{{b}^{2}}}={{\left( \frac{dr}{d\lambda } \right)}^{2}}+\frac{f(r,\alpha ,\beta \text{,}\cdots )}{r^2}.\label{b2}
\end{equation}

We rescale the affine parameter $\lambda$ as $\lambda /\left| L \right|$ , $b=\left| L \right|/E$ is the impact parameter, and $L$ and $E$ are the conserved angular momentum and energy of the photons, respectively. From Eq.(\ref{ll}), the effective potential governing the propagation of light rays is defined as
\begin{equation}
{{V}_{eff}}(r)=\frac{\sqrt{f(r,\alpha ,\beta \text{,}\cdots )}}{r}.\label{veff}
\end{equation}
Unstable circular null geodesics radius ${{r}_{ph}}$ , which form a photon sphere of radius ${{r}_{ph}}$, are determined by the following conditions
\begin{equation}
{{V}_{eff}}({{r}_{ph}})=\frac{1}{{{b}_{ph}}},~~~ V{{'}_{eff}}({{r}_{ph}})=0,~~~ V'{{'}_{eff}}({{r}_{ph}})<0,\label{vcon}
\end{equation}
where ${{b}_{ph}}$ is the corresponding impact parameter.

For a fixed impact parameter, the radial motion of photons can be treated analogously to the motion of a particle in a classical potential field (e.g., gravitational or centrifugal potential). This analogy facilitates the visualization of the photon's dynamical evolution. Here, the radius of the unstable photon sphere is denoted by $r_ph$, which is determined by the implicit expression
\begin{equation}
{{r}_{ph}}f^\prime({{r}_{ph}},\alpha ,\beta ,\cdots )-2f({{r}_{ph}},\alpha ,\beta ,\cdots )=0.\label{rph}
\end{equation}
The critical impact parameter is given by
\begin{equation}
{{b}_{ps}}=\frac{{\bar{L}}}{{{E}_{c}}}={{\left. \frac{r}{\sqrt{f(r,\alpha ,\beta ,\cdots )}} \right|}_{{{r}_{ph}}}}.\label{bph}
\end{equation}

\section{Application of Different Entropy Corrections}\label{application}
\subsection{Barrow Entropy}

Substituting Eq. (\ref{Sbb}) into Eq. (\ref{fr}), we obtain the spacetime metric corresponding to the Barrow entropy
\begin{equation}
{{f}_{B}}(r,\Delta )=\frac{2}{(2+\Delta ){{\pi }^{\Delta /2}}{{r}^{\Delta }}}f(r).\label{fra}
\end{equation}
From Eq. (\ref{fra}), the position of the black hole¡¯s event horizon satisfies $f({{r}_{+}})=0$ , the black hole¡¯s radiation temperature $T=\frac{1}{4\pi }{{\left( \frac{\partial {{S}_{BH}}}{\partial S} \right)}_{r={{r}_{+}}}}f'({{r}_{+}})$, and the black hole¡¯s parameters are $M$. As can be seen from Eq. (\ref{fra}), when $0 < \Delta < 1$ and $ r>2M $, the factor $\frac{2}{(2+\Delta ){{\pi }^{\Delta /2}}{{r}^{\Delta }}} $ is continuous and smooth, which indicates that ${f}_{B}(r,\Delta)$ possesses no horizons other than the event horizon $r=2M$. The Kretschmann scalar

\begin{equation}
R_{\mu\nu\rho\sigma}R^{\mu\nu\rho\sigma}=\frac{4-8 {f}_{B}(r,\Delta )+4 f^{2}_{B}(r,\Delta)+4 r^2 {f}^{' 2}_{B}(r,\Delta )+r^4 {f}^{'' 2}_{B}(r,\Delta )}{r^4}. \label{ruv}
\end{equation}
Outside the event horizon, the curvature quantities remain finite and well-behaved everywhere.

From Eqs. (\ref{gtt}) and (\ref{gbb}), we obtain
\begin{eqnarray}
&G_{t}^{t}&=G_{r}^{r}=\frac{2}{(2+\Delta ){{\pi }^{\Delta /2}}{{r}^{\Delta +2}}}\left( \frac{2M\Delta }{r}-\Delta +1 \right)-\frac{1}{{{r}^{2}}}, \nonumber\\
&G_{t}^{t}&=G_{r}^{r}=\frac{2\Delta }{(2+\Delta ){{\pi }^{\Delta /2}}{{r}^{\Delta +2}}}\left( \frac{2M-r}{r} \right)+\frac{1}{{{r}^{2+\Delta }}}\left( \frac{2}{(2+\Delta ){{\pi }^{\Delta /2}}}-{{r}^{\Delta }} \right),\nonumber\\
&G_{\varphi }^{\varphi }&=G_{\theta }^{\theta }=\frac{2}{(2+\Delta ){{\pi }^{\Delta /2}}{{r}^{\Delta +2}}}\left( \frac{2M\Delta }{r}-\Delta +1 \right)-\frac{1}{{{r}^{2}}}. \label{gtr}
\end{eqnarray}

For $0\le \Delta \le 1$, as $r\geq 2M$, $\frac{2\Delta }{(2+\Delta ){{\pi }^{\Delta /2}}{{r}^{\Delta +2}}}\left( \frac{2M-r}{r} \right)\leq0$, taking $M=1$ gives $r\geq2$, hence $\left( \frac{2}{(2+\Delta ){{\pi }^{\Delta /2}}}-{{r}^{\Delta }} \right)\leq0$. When $\Delta =0$, the metric reduces to the Schwarzschild vacuum.
 We now briefly examine the energy conditions of the corrected spacetime. For $M>0$ and $\Delta >0$, the effective energy density is positive $\rho >0$. Since $\rho +{{p}_{r}}=0$ holds along the radial null direction, the null energy condition (NEC) is saturated. Along the tangential direction, we obtain $\rho +{{p}_{t}}=0$ , which also leads to saturation of the NEC. Furthermore, $\rho >0$ ensures that the weak energy condition (WEC) is satisfied. We find that $\rho +{{p}_{r}}+2{{p}_{t}}=2p\le 0$, implying that the strong energy condition (SEC) is violated. The dominant energy condition (DEC) holds only when $\rho >0$ and $\left| {{p}_{i}} \right|\le \rho $. As observed, the effective matter sector behaves as an isotropic fluid with a positive energy density equal to the radial pressure. Such stress-energy configurations cannot be produced by conventional classical matter and originate from the quantum-gravitational or fractal corrections introduced by the Barrow entropy modification. Therefore, the SEC violation is not a pathological feature of the spacetime; instead, it indicates the existence of an effective non-classical matter source that sustains the modified horizon geometry \cite{AA-2026b}.

Substituting Eq. (\ref{fra}) into Eq. (\ref{veff}), we obtain
\begin{equation}
{{V}_{eff}}(r)=\frac{1}{r}\sqrt{\frac{2}{(2+\Delta ){{\pi }^{\Delta /2}}{{r}^{\Delta }}}f(r)}.\label{veffa}
\end{equation}
Taking $r=M\eta$ ($2<\eta <\infty$), we have
\begin{equation}
{{(M)}^{1+\Delta /4}}{{V}_{eff}}(\eta )=\frac{1}{{{\eta }^{1+\Delta /4}}}\sqrt{\frac{2}{(2+\Delta ){{\pi }^{\Delta /2}}}\left( 1-\frac{2}{\eta } \right)}.\label{mveffa}
\end{equation}
As $\Delta$ with different values, we obtain the corresponding ${{(2M)}^{1+\Delta /4}}{{V}_{eff}}(\eta )-\eta $ curves based on Eq. (\ref{mveffa}).
\begin{figure}[!htpb]
 \includegraphics[width=4.5cm,height=3.5cm]{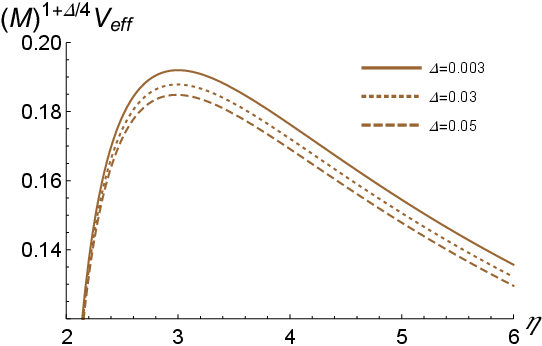}
 \caption{The ${M}^{1+\Delta/4}V_{eff}(\eta)-\eta$ curves corresponding to different values of $\Delta$. }\label{Fig.vnam}
\end{figure}

Substituting Eq. (\ref{fra}) into Eqs. (\ref{rph}) and (\ref{bph}), we get

\begin{eqnarray}
&{{r}_{ph\Delta }}&=\frac{2M(3+\Delta )}{2+\Delta }, ~~~~~~                         \nonumber\\
&{{b}_{ps\Delta }}&=\frac{\sqrt{2}M{{(3+\Delta )}^{3/2}}{{\pi }^{\Delta /4}}}{{{(2+\Delta )}^{1/2}}}{{\left( \frac{2M(3+\Delta )}{2+\Delta } \right)}^{\Delta /2}}.\label{rba}
\end{eqnarray}
Here ${r}_{ph\Delta }$ and ${b}_{ps\Delta }$ stand for the photon sphere radius and corresponding critical impact parameter of the black hole with Barrow entropy.
\begin{figure}[!htp]
 \includegraphics[width=4.5cm,height=3.5cm]{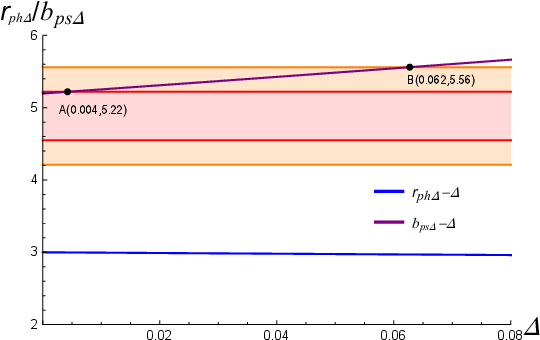}
 \caption{the ${{r}_{ph\Delta }}-\Delta $ curve and ${{b}_{ps\Delta }}-\Delta $ curve when $M=1$, $0\le \Delta \le 1$. }\label{Fig.rba}
\end{figure}

Figure \ref{Fig.rba} presents the ${{r}_{ph\Delta }}-\Delta $ curve and ${{b}_{ps\Delta }}-\Delta $ curve when $M=1$, $0\le \Delta \le 1$. It can be seen from Fig. \ref{Fig.vnam} and Fig. \ref{Fig.rba} that with the increase of the entropy correction parameter $\Delta$, the maximum value of the effective potential  decreases, the position of the photon sphere ${{r}_{ph\Delta }}$ decreases while the critical impact parameter ${{b}_{ph\Delta }}$ increases.

The latest Very Long Baseline Interferometry (VLBI) observational results of Sagittarius A* from the Event Horizon Telescope (EHT) \cite{ RCR-2026,SV-2023,RCP-2023,XXZ-2025e} are given by
\begin{eqnarray}
4.55\le {{b}_{ps\alpha }}\le 5.22,~~~ 1.516 \le \frac{{b}_{ps\alpha }}{3} \le 1.74,~~~( 1\sigma ),\nonumber \\
4.21\le {{b}_{ps\alpha }}\le 5.56,~~~ 1.403 \le \frac{{b}_{ps\alpha }}{3} \le 1.85,~~~( 2\sigma ). \label{con}
\end{eqnarray}

Constrained by Eq. (\ref{con}), the range of the parameter $\Delta$  in the Barrow entropy is determined as $0\le \Delta \le 0.004$ $(\dot{}1\sigma )$  (corresponding to point $A$ in Fig. \ref{Fig.rba} and $0\le \Delta \le 0.062$ (corresponding to point $B$ in Fig. \ref{Fig.rba}). The region between the two red lines is the constraint interval of $(1\sigma )$ , and the region between the two orange lines is the constraint interval of $(2\sigma )$.
\subsection{R$\acute{e}$nyi Entropy}

Substituting Eq. (\ref{Srb}) into Eq. (\ref{fr}), we obtain the spacetime metric corresponding to the R$\acute{e}$nyi entropy
\begin{equation}
{{f}_{R}}(r,\lambda )=(1+\pi \lambda {{r}^{2}})f(r).\label{frr}
\end{equation}
As can be seen from Eq. (\ref{frr}), when $0<\lambda<1$ and $ r>2M $, the factor $(1+\pi \lambda {{r}^{2}}) $ is continuous and smooth, which indicates that ${f}_{R}(r,\lambda )$ possesses no horizons other than the event horizon $r=2M$. In this case, the Kretschmann scalar has the same form with Eq. (\ref{ruv}), analysis indicates it has well-behaved behavior with no divergence.

Using Eqs. (\ref{gtt}) and (\ref{gbb}), we derive
\begin{eqnarray}
&G_{t}^{t}&=G_{r}^{r}=3\pi \lambda -4\pi \lambda \frac{M}{r}=\pi \lambda \left( 3-\frac{4M}{r} \right), \nonumber \\
&G_{\theta }^{\theta }&=\pi \lambda \left( 3-\frac{2M}{r} \right). \label{gttr}
\end{eqnarray}
When $\lambda =0$ , the metric reduces to the Schwarzschild vacuum. Note that for $M>0$,  $r>2M$, the effective energy density is $\rho <0$. Since $\rho +{{p}_{r}}=0$ along the radial null direction, the NEC is saturated. Along the tangential direction we have $\rho +{{p}_{t}}=\frac{2M}{r}>0$, so the NEC is satisfied. Since $\rho <0$, the WEC is violated. We find that $\rho +{{p}_{r}}+2{{p}_{t}}=2p\ge 0$, the SEC holds. Finally, as $\rho \ge 0$ and $\left| {{p}_{i}} \right|\le \rho $ fail to hold, the DEC is also violated.

Substituting Eq. (\ref{frr}) into Eq. (\ref{veff}), we obtain
\begin{equation}
{{V}_{eff}}(r)=\frac{1}{r}\sqrt{(1+\pi \lambda {{r}^{2}})f(r)},\label{veffr}
\end{equation}
Taking $r=3M\eta$ ($\frac{2}{3}<\eta<\infty$), we have
\begin{equation}
M{{V}_{eff}}(\eta)=\frac{\sqrt{(1+3\chi \frac{{{\eta }^{2}}}{4})\left( 1-\frac{2}{3\eta } \right)}}{3\eta },\label{mveffr}
\end{equation}
where $\chi =12\pi \lambda {{M}^{2}}$.
As $\chi$ with different values, we obtain the corresponding ${M}{{V}_{eff}}-\eta $ curves based on Eq. (\ref{mveffr}).
\begin{figure}[!htp]
 \includegraphics[width=4.5cm,height=3.5cm]{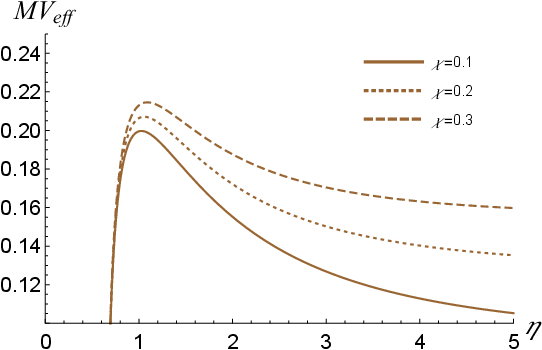}
 \caption{The ${M}V_{eff}-\eta$ curves corresponding to different values of $\chi$. }\label{Fig.vdx}
\end{figure}

\begin{figure}[!htp]
 \includegraphics[width=4.5cm,height=3.5cm]{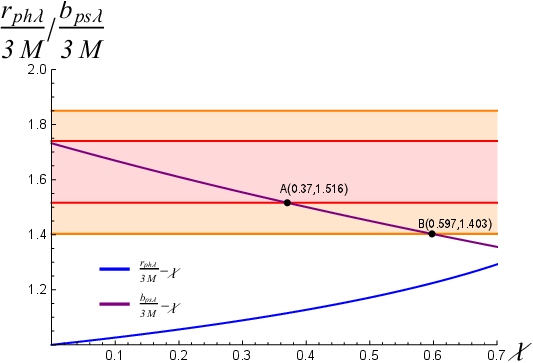}
 \caption{the $\frac{{r}_{ph\lambda }}{3M}-\chi $ curve and $\frac{{b}_{ps\lambda }}{3M}-\chi $ curve when $M=1$. }\label{Fig.rbr}
\end{figure}

Substituting Eq. (\ref{frr}) into Eqs. (\ref{rph}) and (\ref{bph}), we get
\begin{eqnarray}
&\frac{{{r}_{ph\lambda }}}{3M}&=2\frac{1-\sqrt{1-\chi }}{\chi },   \nonumber \\
&\frac{{{b}_{ps\lambda }}}{3M}&=\frac{2(1-\sqrt{1-\chi })}{\chi \sqrt{\left( 1-\frac{\chi }{3(1-\sqrt{1-\chi })} \right)\left( 1+\tfrac{3}{\chi }{{\left( (1-\sqrt{1-\chi }) \right)}^{2}} \right)}}.\label{rbr}
\end{eqnarray}
Here ${r}_{ph\lambda }$ and ${b}_{ps\lambda }$ are the photon sphere radius and corresponding critical impact parameter of the black hole with the R$\acute{e}$nyi entropy.
Figure \ref{Fig.rbr} presents the $\frac{{r}_{ph\lambda }}{3M}-\chi $ curve and $\frac{{b}_{ps\lambda }}{3M}-\chi $ curve when $M=1$.
It can be seen from Fig. \ref{Fig.vdx} and Fig. \ref{Fig.rbr} that with the increase of the entropy correction parameter $\chi$ , the maximum value of the potential ${{V}_{eff}}(\eta )$ increases, the position of the photon sphere $\frac{{{r}_{ph\lambda }}}{3M}$ increases, the critical impact parameter $\frac{{{b}_{ps\lambda }}}{3M}$ decreases. Constrained by Eq. (\ref{con}), the value range of the parameter $\chi (\lambda )$ in the R$\acute{e}$nyi entropy is determined as $0\le \chi (\lambda )\le 0.37$, $(1\sigma )$ (corresponding to point $A$ in Fig. \ref{Fig.rbr}) and $0\le \chi (\lambda )\le 0.597$, $(2\sigma )$ (corresponding to point $B$ in Fig. \ref{Fig.rbr}).

Comparing Fig. \ref{Fig.rba} with Fig. \ref{Fig.rbr}, it is found that the variation laws of the photon sphere position ${r}_{ph}$ and the critical impact parameter ${b}_{ps}$ with the parameter $\Delta$ in the Barrow entropy are different from those with the parameter $\lambda$ in the R$\acute{e}$nyi entropy, and the two are exactly opposite. From Fig. \ref{Fig.vnam} and Fig. \ref{Fig.vdx}, the variation laws of the effective potential ${V}_{eff}$ with $\Delta$ and $\lambda$ are also exactly opposite. This indicates that the parameter $\Delta$  in the Barrow entropy and the parameter $\lambda$ in the R$\acute{e}$nyi entropy play different roles in the black hole deformation.
The spacetime in Eq. (\ref{frr}) is not asymptotically flat. In such spacetimes, although the critical impact parameter $b_{ps}$ given in Eq. (\ref{bph}) can be defined globally, its direct
correspondence to the black hole¡¯s shadow size viewed by an observer is no longer straightforward. A more precise discussion would require the application of the ray-tracing method. However, for the sake of simplicity, this work only discusses the critical impact parameter $b_{ps}$ , and a more quantitative ray-tracing analysis is not taken into consideration.
\subsection{Sharma-Mittal entropy}
 Substituting Eq. (\ref{Sbs}) into Eq. (\ref{fr}), we obtain the spacetime metric corresponding to the Sharma-Mittal entropy
\begin{equation}
{{f}_{SM}}(r,\alpha ,\beta )=\frac{1}{{{(1+\pi \beta {{r}^{2}})}^{\tfrac{\alpha }{\beta }-1}}}f(r).\label{frs}
\end{equation}
 As can be seen from Eq. (\ref{frs}), when $\alpha >0$, $\beta>0$, the factor $\frac{1}{{{(1+\pi \beta {{r}^{2}})}^{\tfrac{\alpha }{\beta }-1}}}$ is continuous and smooth for $ r>2M $, which indicates that ${f}_{SM}(r,\alpha ,\beta )$ possesses no horizons other than the event horizon $r=2M$. In this case, the Kretschmann scalar has the same form with Eq. (\ref{ruv}), analysis indicates it has well-behaved behavior with no divergence.
Using Eqs. (\ref{gtt}) and (\ref{gbb}), we derive

\begin{eqnarray}
&G_{t}^{t}&=G_{r}^{r}=\frac{1}{{{(1+\pi \beta {{r}^{2}})}^{\tfrac{\alpha }{\beta }}}}\left( 2\pi (\beta -\alpha )f(r)+\frac{(1+\pi \beta {{r}^{2}})}{{{r}^{2}}} \right)-\frac{1}{{{r}^{2}}},\nonumber\\
&G_{\varphi }^{\varphi }&=G_{\theta }^{\theta }=\frac{\pi (\beta -\alpha )}{{{(1+\pi \beta {{r}^{2}})}^{\tfrac{\alpha }{\beta }+1}}}\left[ -2\pi \alpha {{r}^{2}}f(r)+(1+\pi \beta {{r}^{2}})f(r)+2(1+\pi \beta {{r}^{2}}) \right].\label{gtts}
\end{eqnarray}

When $\alpha \rightarrow \beta$, $G_{t}^{t}=G_{r}^{r}\to 0$, $G_{\theta }^{\theta }\to 0$, the metric reduces to the Schwarzschild vacuum. When $\alpha \to 0$, $\beta \ge 0$, we have
\begin{eqnarray}
&G_{t}^{t}&=G_{r}^{r}=2\pi \beta f(r)+\pi \beta =\pi \beta \left( 3-\frac{4M}{r} \right)\ge 0,~~~~ \rho \le 0\nonumber\\
&G_{\varphi }^{\varphi }&=G_{\theta }^{\theta }=\pi \beta \left[ f(r)+2 \right]\ge 0. \label{gtrs}
\end{eqnarray}
Note that for $M>0$, $\alpha \ge 0$, $\beta >0$ and $r>2M$, the effective energy density satisfies $\rho <0$. Since $\rho +{{p}_{r}}=0$ along the radial null direction, the NEC is saturated. Along the tangential direction we have $\rho +{{p}_{t}}=\frac{4M}{r}>0$, so the NEC holds. As $\rho <0$, the WEC is violated. We obtain $\rho +{{p}_{r}}+2{{p}_{t}}=\pi \beta \left[ f(r)+2 \right]\ge 0$, which implies the SEC is satisfied. Finally, the requirement $\rho \ge 0$ and $\left| {{p}_{i}} \right|\le \rho $ cannot be fulfilled, so the DEC is also violated.
For $\beta >\alpha >0$,
\begin{eqnarray}
&G_{t}^{t}&=G_{r}^{r}=\frac{2\pi (\beta -\alpha )}{{{(1+\pi \beta {{r}^{2}})}^{\tfrac{\alpha }{\beta }}}}f(r)+\frac{{{(1+\pi \beta {{r}^{2}})}^{1-\tfrac{\alpha }{\beta }}}}{{{r}^{2}}}-\frac{1}{{{r}^{2}}}>0,~~~~\rho \le 0, \nonumber\\
&G_{\varphi }^{\varphi }&=G_{\theta }^{\theta }>0.
\end{eqnarray}
Accordingly, the set of valid energy conditions for $\beta >\alpha $ is identical to that in the limit $\alpha \to 0$ with $\beta \ge 0$. The SEC remains satisfied, whereas the DEC is violated because the constraints $\rho \ge 0$ and $\left| {{p}_{i}} \right|\le \rho $ fail to hold.

Substituting Eq. (\ref{frs}) into Eqs. (\ref{rph}) and (\ref{bph}), we obtain that the position of the photon sphere satisfies
\begin{equation}
3M+M\pi r_{ph}^{2}(2\alpha +\beta )-{{r}_{ph}}-\alpha \pi r_{ph}^{3}=0,\label{rcon}
\end{equation}
Introducing dimensionless parameters $x=\alpha \pi {{M}^{2}}$, $y=\beta \pi {{M}^{2}}$, and ${{r}_{phxy}}={r}_{ph}=3M\eta$ ($\frac{2}{3}<\eta<\infty$), Eq. (\ref{rcon}) can be transformed into
\begin{equation}
1+3{{\eta }^{2}}\left( 2x+y \right)-\eta -9x{{\eta }^{3}}=0,\label{conxy}
\end{equation}
When $x(\alpha) = 0$, $y(\beta)= 0$,  Eq. (\ref{conxy}) reduces to the photon sphere position of the Schwarzschild black hole.
\\

In the following we will solve Eq. (\ref{conxy}) by considering two cases.

\textbf{1) As $x(\alpha) = 0$.}

\begin{figure}[!htp]
 \includegraphics[width=4.5cm,height=3.5cm]{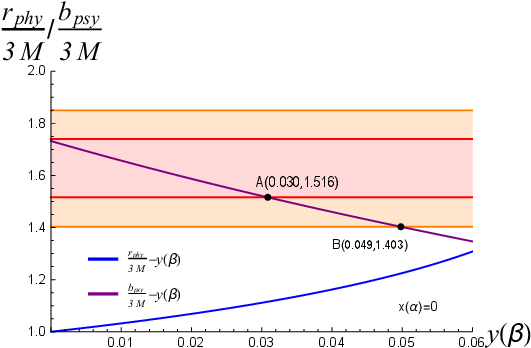}
 \caption{the $\frac{{r}_{phy }}{3M}-y(\beta) $ curve and $\frac{{b}_{psy }}{3M}-y(\beta) $ curve when $M=1$. }\label{Fig.rbs1}
\end{figure}

\begin{figure}[!htp]
 \includegraphics[width=4.5cm,height=3.5cm]{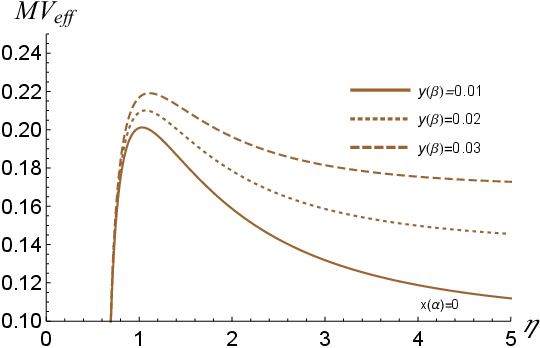}
 \caption{The ${M}V_{eff}-\eta$ curves corresponding to different values of $y(\beta)$. }\label{Fig.vnyx0}
\end{figure}

Eq. (\ref{conxy}) is transformed into
\begin{equation}
1+3y{{\eta }^{2}}-\eta =0.\label{cony}
\end{equation}
Solving this equation, we obtain $\eta =\frac{1\pm \sqrt{1-12y}}{6y}$. It can be inferred from this that $0\le y\le \frac{1}{12}$, and only the negative sign can be taken. This is because when the positive sign is adopted, $y\rightarrow 0$, $\eta$ diverges, which does not satisfy the requirement. Now we get
\begin{eqnarray}
&\eta& =\frac{1-\sqrt{1-12y}}{6y},~~~~~~~  \frac{{{r}_{phy}}}{3M}=\frac{1-\sqrt{1-12y}}{6y}, \nonumber\\
&\frac{{{b}_{psy}}}{3M}&=\frac{{\bar{L}}}{{{E}_{c}}}={{\left. \frac{r}{\sqrt{f(r,y)}} \right|}_{{{r}_{phy}}}}=\frac{1-\sqrt{1-12y}}{6y\sqrt{\left( 1+\frac{1}{4y}{{(1-\sqrt{1-12y})}^{2}} \right)\left( 1-\frac{4y}{1-\sqrt{1-12y}} \right)}}. \label{rbsy}
\end{eqnarray}
Note that in this case ${r}_{phy}={r}_{ph}$ and ${b}_{psy}={b}_{ps}$, stand for the photon sphere radius and corresponding critical impact parameter of the black hole with the Sharma-Mittal entropy entropy as $x(\alpha)=0 $.
Figure \ref{Fig.rbs1} presents the $\frac{{r}_{phy }}{3M}-y(\beta) $ curve and $\frac{{b}_{psy }}{3M}-y(\beta) $ curve when $M=1$, $x(\alpha)=0$.
It can be seen from Fig. \ref{Fig.vnyx0} and Fig. \ref{Fig.rbs1} that when $x(\alpha) = 0$, the maximum value of the effective potential ${V}_{eff}$ increases with the increase of $y(\beta)$. For the black hole corresponding to the Sharma-Mittal entropy, the position of the photon sphere ${r}_{phy }$ increases with the increase of $y(\beta)$, while the critical impact parameter ${b}_{psy }$  decreases. Constrained by Eq. (\ref{con}), when $x(\alpha) = 0$, the value range of the parameter $y(\beta)$ in the Sharma-Mittal entropy is determined as $0\le y(\beta )\le 0.030$, $(1\sigma )$  (corresponding to point $A$ in Fig. \ref{Fig.rbs1}) and $0\le y(\beta )\le 0.049$, $(2\sigma )$ (corresponding to point $B$ in Fig. \ref{Fig.rbs1}).

Substituting Eq. (\ref{frs}) into Eq. (\ref{veff}), we obtain
\begin{equation}
M{{V}_{eff}}(\eta)=\frac{1}{3\eta }\sqrt{\frac{1}{{{(1+9y{{\eta }^{2}})}^{\frac{x}{y}-1}}}\left( 1-\frac{2}{3\eta } \right)},\label{mveffs}
\end{equation}
when $x(\alpha) =0$,
\begin{equation}
M{{V}_{eff}}(\eta)=\frac{1}{3\eta }\sqrt{(1+9y{{\eta }^{2}})\left( 1-\frac{2}{3\eta } \right)}.\label{mveffs0}
\end{equation}
As $y(\beta) $ with different values, we obtain the corresponding ${M}{{V}_{eff}}-\eta $ curves based on Eq. (\ref{mveffs0}).

Comparing Fig. \ref{Fig.rbr} with Fig. \ref{Fig.rbs1}, it is found that the variation laws of the photon sphere position ${r}_{ph\lambda }$ for the black hole corresponding to the R$\acute{e}$nyi entropy, and the photon sphere position ${r}_{phy }$ and critical impact parameter ${b}_{psy }$ for the black hole corresponding to the Sharma-Mittal entropy when $x(\alpha) = 0$, with respect to the correction parameter $y(\beta)$ are similar. Comparing Fig. \ref{Fig.vnyx0} with Fig. \ref{Fig.vdx}, the variation laws of the effective potential ${V}_{eff}$ after entropy correction with respect to $r(\eta)$ are also similar. Thus, the optical properties of the black hole corresponding to the R$\acute{e}$nyi entropy are similar to those of the black hole corresponding to the Sharma-Mittal entropy when $x(\alpha) = 0$.

\textbf{2) As $y(\beta)=constant\neq0$ or $x(\alpha)=constant\neq0$.}

The solution to Eq. (\ref{conxy}) is
\begin{equation}
\eta =Y+\frac{3\text{(2}x+y\text{)}}{27x}=Y+\frac{\text{(2}x+y\text{)}}{9x},\label{conxys}
\end{equation}
where
\begin{eqnarray}
&Y&=\sqrt[3]{-\frac{q}{2}+\sqrt{{{\left( \frac{q}{2} \right)}^{2}}+{{\left( \frac{p}{3} \right)}^{3}}}}+\sqrt[3]{-\frac{q}{2}-\sqrt{{{\left( \frac{q}{2} \right)}^{2}}+{{\left( \frac{p}{3} \right)}^{3}}}},\nonumber \\
&p&=-\frac{1}{9x}\left( \frac{{{(2x+y)}^{2}}}{3x}-1 \right), ~~~~~ q=-\frac{1}{9x}\left( 1+2\frac{{{(2x+y)}^{3}}}{{{(9x)}^{2}}}-\frac{\text{(2}x+y\text{)}}{9x} \right).\label{ypq}
\end{eqnarray}
Substituting Eqs. (\ref{conxys}) and (\ref{ypq})into Eqs. (\ref{veff}), (\ref{rph}) and (\ref{bph}), we can obtain
\begin{eqnarray}
&M{{V}_{eff}}(r,x,y)&=\frac{1}{3\eta }\sqrt{{{(1+9y{{\eta }^{2}})}^{1-\tfrac{x}{y}}}\left( 1-\frac{2}{3\eta } \right)} ,\label{mveffs2}\\
&\frac{{{r}_{phxy}}}{3M}=\eta,&~~~\frac{{{b}_{psxy}}}{3M}=\frac{{\bar{L}}}{{{E}_{c}}}={{\left. \frac{r}{\sqrt{f(r,x,y)}} \right|}_{{{r}_{phxy}}}}=\frac{\eta }{\sqrt{{{(1+9y{{\eta }^{2}})}^{1-\tfrac{x}{y}}}\left( 1-\frac{2}{3\eta } \right)}}. \label{rbsxy}
\end{eqnarray}
Note that in this case, ${r}_{phxy}={r}_{ph}$, ${b}_{psxy}={b}_{ps}$ stand for the photon sphere radius and corresponding impact parameter, respectively.  In the following, ${r}_{phxy}={r}_{ph}={r}_{phx}$, ${b}_{psxy}={b}_{ps}={b}_{psx}$ for a fixed $y(\beta )$, and ${r}_{phxy}={r}_{ph}={r}_{phy}$, ${b}_{psxy}={b}_{ps}={b}_{psy}$ for a fixed $x(\alpha )$.
From Eqs. (\ref{conxys}) and (\ref{rbsxy}), we can plot the $\frac{{{b}_{psx}}}{3M}-x(\alpha )$  curve and $\frac{{{r}_{phx}}}{3M}-x(\alpha )$ for a given $y(\beta)$ , the $\frac{{{b}_{psy}}}{3M}-y(\beta )$  curve and $\frac{{{r}_{phy}}}{3M}-x(\alpha )$  for a given $x(\alpha )$. Given the constraint conditions satisfied by $x(\alpha )$ and $y(\beta )$, we have also plotted the effective potential $M{{V}_{eff}}(r,x,y)-\eta $ curves for fixed $x(\alpha )$ or $y(\beta )$ based on the Eq. (\ref{mveffs2}).

\begin{figure*}[!htb]
\subfigure[]{\includegraphics[width=5cm,height=4.5cm]{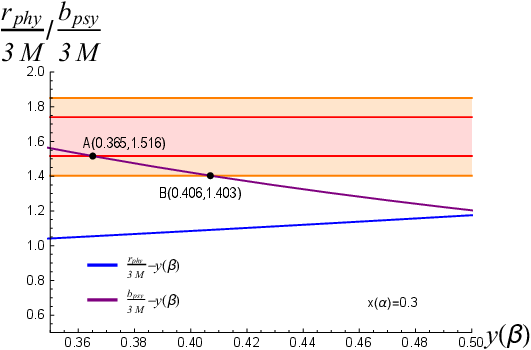}}~~
\subfigure[]{\includegraphics[width=5cm,height=4.5cm]{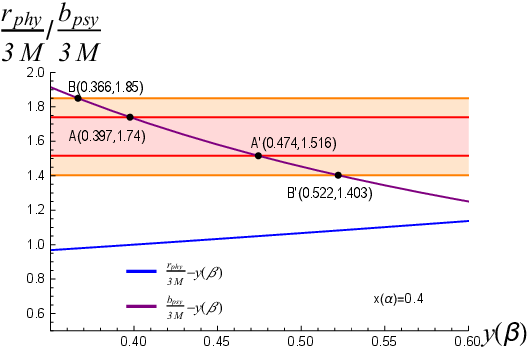}}~~
\subfigure[]{\includegraphics[width=5cm,height=4.5cm]{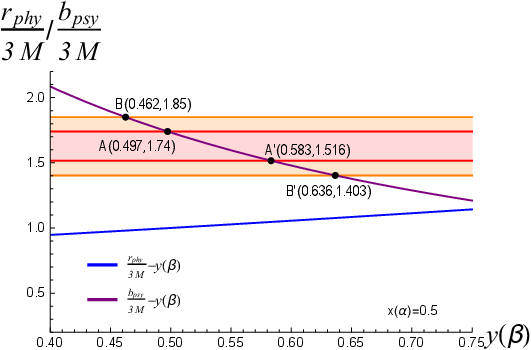}}
\vskip -1mm \caption{the $\frac{{r}_{phy }}{3M}-y(\beta) $ curve and $\frac{{b}_{psy }}{3M}-y(\beta) $ curve with different values of $x(\alpha )$.}\label{Fig.rbsx}
\end{figure*}
\begin{figure}[!htp]
 \includegraphics[width=4.5cm,height=3.5cm]{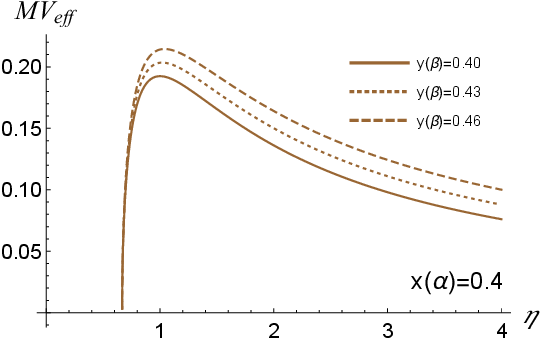}
 \caption{The ${M}V_{eff}-\eta$ curves corresponding to different values of $y(\beta)$ when $x(\alpha )=0.4$. }\label{Fig.vnsy}
\end{figure}
\begin{figure*}[!htb]
\subfigure[]{\includegraphics[width=5cm,height=4.5cm]{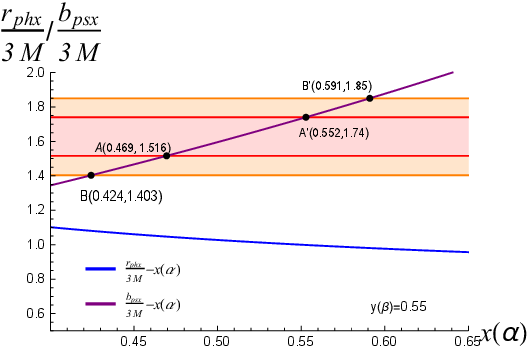}}~~
\subfigure[]{\includegraphics[width=5cm,height=4.5cm]{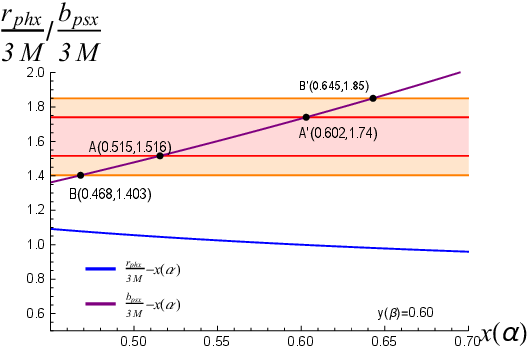}}~~
\subfigure[]{\includegraphics[width=5cm,height=4.5cm]{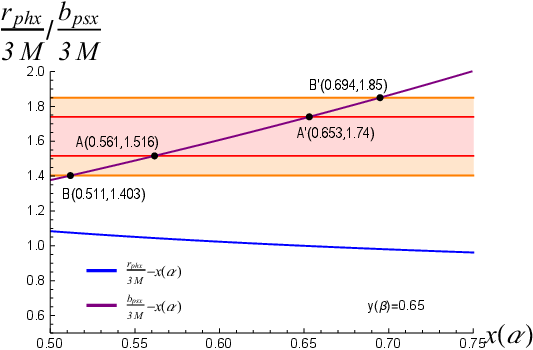}}
\vskip -1mm \caption{the $\frac{{r}_{phy }}{3M}-x(\alpha) $ curve and $\frac{{b}_{psy }}{3M}-x(\alpha) $ curve with different values of $y(\beta )$.}\label{Fig.rbsy}
\end{figure*}
\begin{figure}[!htp]
 \includegraphics[width=4.5cm,height=3.5cm]{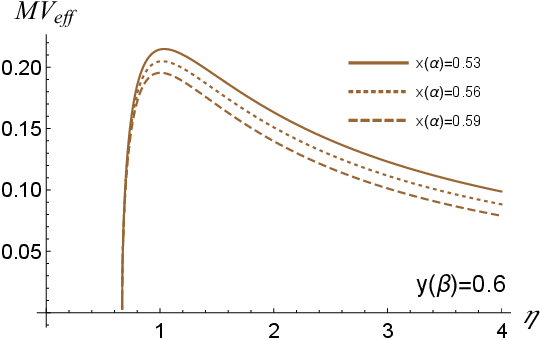}
 \caption{The ${M}V_{eff}-\eta$ curves corresponding to different values of $x(\alpha )$ when $y(\beta )=0.6$. }\label{Fig.vnsx}
\end{figure}
It can be seen from Fig. \ref{Fig.rbsx} and Fig. \ref{Fig.vnsy} that for a fixed value of $x(\alpha)$, the photon sphere radius ${r}_{phy }$ of the black hole associated with Sharma-Mittal entropy increases as the correction parameter $y(\beta)$ rises, whereas the critical impact parameter ${b}_{psy }$ decreases accordingly. Meanwhile, the maximum value of the effective potential $V_{eff}$ gets enlarged with increasing $y(\beta)$. In Fig. \ref{Fig.rbsx}, we have marked the allowable ranges of $y(\beta)$ constrained by Eq. (\ref{con}) under different values of $x(\alpha)$. It is evident that the bounded interval of expands toward larger values as $x(\alpha)$ increases. Constrained by Eq. (\ref{con}), when $x(\alpha)=0.3$, the permissible range of parameter $y(\beta)$ in the Sharma-Mittal entropy is $0\le y(\beta )\le 0.365$, $(1\sigma )$ given by (corresponding to point $A$ in Fig. \ref{Fig.rbsx}(a)) and $0\le y(\beta )\le 0.406$, $(2\sigma )$ (corresponding to point $B$ in Fig. \ref{Fig.rbsx}(a)). When $x(\alpha)=0.4$, the corresponding range of is $0.397\le y(\beta )\le 0.474$, $(1\sigma )$ (corresponding to points $A$ and $A^\prime$ in Fig. \ref{Fig.rbsx}(b)) and $0.366\le y(\beta )\le 0.522$, $(2\sigma )$ (corresponding to points $B$ and $B^\prime$ in Fig. \ref{Fig.rbsx}(b) ). When $x(\alpha)=0.5$, the allowed values of are $0.497\le y(\beta )\le 0.583$, $(1\sigma )$ (corresponding to points $A$ and $A^\prime$ in Fig. \ref{Fig.rbsx}(c)) and $0.\text{462}\le y(\beta )\le 0.636$, $(2\sigma )$ (corresponding to points $B$ and $B^\prime$ in Fig. \ref{Fig.rbsx}(c)).

It can be seen from Fig. \ref{Fig.rbsy} and Fig. \ref{Fig.vnsx} that for a fixed value of $y(\beta)$, the photon sphere radius ${r}_{phx }$ of the black hole associated with Sharma-Mittal entropy decreases as the correction parameter $x(\alpha)$ increases, while the critical impact parameter ${b}_{psx }$  grows accordingly. In addition, the maximum value of the effective potential $V_{eff}$ decreases with the increasing $x(\alpha)$.  Constrained by Eq. (\ref{con}), when $y(\beta)=0.55$, the range of the parameter $x(\alpha)$ for Sharma-Mittal entropy is confined to $0.452\le x(\alpha )\le 0.469$, $(1\sigma )$, (corresponding to points $A$ and $A^\prime$ in Fig. \ref{Fig.rbsy}(a)) and $0.424\le x(\alpha )\le 0.519$, $(2\sigma )$, (corresponding to points $B$ and $B^\prime$ in Fig. \ref{Fig.rbsy}(a)). When $y(\beta)=0.60$, the admissible parameter range is $0.515\le x(\alpha )\le 0.602$, $(1\sigma )$ (corresponding to points $A$ and $A^\prime$ in Fig. \ref{Fig.rbsy}(b)) and $0.468\le x(\alpha )\le 0.642$, $(2\sigma )$ (corresponding to points $B$ and $B^\prime$ in Fig. \ref{Fig.rbsy}(b)). when $y(\beta)=0.65$, the range of the parameter $x(\alpha)$ for Sharma-Mittal entropy is confined to $0.561\le x(\alpha )\le 0.653$, $(1\sigma )$, (corresponding to points $A$ and $A^\prime$ in Fig. \ref{Fig.rbsy}(c)) and $0.511\le x(\alpha )\le 0.694$, $(2\sigma )$, (corresponding to points $B$ and $B^\prime$ in Fig. \ref{Fig.rbsy}(c)).
\section{conclusion}\label{conslusion}

By investigating the photon sphere radius, critical impact parameter, and effective potential in spacetimes modified by different entropy corrections, and comparing the numerical results with the latest Event Horizon Telescope (EHT) observations of Sagittarius A*, we constrain the valid ranges of the characteristic parameters introduced in various entropy correction models. Accordingly, our findings provide a meaningful reference for further exploring the parameter constraints of modified entropies.
In this work, we demonstrate that distinct modified entropies lead to completely different physical results. For the Barrow entropy, an increase of the correction parameter $\Delta$ reduces the photon sphere radius ${r}_{ph\Delta }$, raises the critical impact parameter ${b}_{ps\Delta }$, and lowers the effective potential ${V}_{eff}(\eta)$. For the R$\acute{e}$nyi entropy, a larger entropy correction parameter $\lambda$ increases the photon sphere radius ${r}_{ph\lambda }$, decreases the critical impact parameter ${b}_{ps\lambda }$, and strengthens the effective potential ${V}_{eff}(r,\lambda)$.
As illustrated in Fig. \ref{Fig.rbsx} and Fig. \ref{Fig.vnsy}, for a fixed value of $x(\alpha)$, the photon sphere radius ${r}_{phy }$ of the black hole associated with Sharma-Mittal entropy increases as the correction parameter $y(\beta)$ rises, whereas the critical impact parameter ${b}_{psy }$ decreases accordingly, accompanied by a increase of the maximum value of the effective potential $V_{eff}(r,x)$.  By contrast, as shown in Fig. \ref{Fig.rbsy} and Fig. \ref{Fig.vnsx}, with another fixed parameter $y(\beta)$ of the Sharma-Mittal entropy, an increasing correction parameter $x(\alpha)$ causes a smaller photon sphere radius ${r}_{phx }$ , a larger critical impact parameter ${b}_{psx }$, and a lower effective potential $V_{eff}(r,y)$. These results indicate that the evolutionary behaviors of the effective potential, photon sphere radius and critical impact parameter with respect to the entropy correction parameters differ significantly among different modified entropy scenarios.

This conclusion offers valuable insights for revealing the essential physical nature of entropy corrections. Furthermore, the present study opens several promising avenues for future research. First, the analysis can be extended to multi-parameter and rotating black hole spacetimes, to explore the optical properties of entropy-corrected black holes and enrich the research scope of black hole physics. Second, with the continuous improvement of resolution and sensitivity of future Very Long Baseline Interferometry (VLBI) facilities, more precise constraints on the parameters of various entropy corrections can be achieved, which will further optimize relevant theoretical models. Meanwhile, our theoretical results provide timely data comparisons for increasingly accurate astrophysical observations.

\acknowledgments
This work was supported by the National Natural Science Foundation of China ( Grant No. 12375050 ),  the Natural Science Foundation of Shanxi Province (Grant No. 202303021211180, 202203021221209, 202203021221211, 202503021211241), the Program for the InnovativeTalents of Higher Education Institutions of Shanxi (Grant No2024Q030), and the Doctoral Sustentation Fund of Shanxi Datong University.

\end{document}